\documentclass[pra,showpacs,nofootinbib,twocolumn]{revtex4}
\usepackage{bm,dcolumn,amsmath,graphicx}

\begin{document}

\title{Atomic properties of  superheavy elements No, Lr, and Rf}

\author{V. A. Dzuba}
\affiliation{School of Physics, University of New South Wales,
        Sydney, 2052, Australia}

\author{M. S. Safronova}
\affiliation {Department of Physics and Astronomy, University of
  Delaware, Newark, Delaware, USA and \\Joint Quantum Institute, 
NIST and the University of Maryland, College Park, Maryland, USA}

\author{U. I. Safronova}
\affiliation{Physics Department, University of Nevada, Reno, Nevada, USA}

\begin{abstract}

The combination of the configuration interaction method and all-order
single-double coupled-cluster technique is used to calculate excitation
energies, ionization potentials and static dipole
polarizabilities of superheavy elements nobelium, lawrencium and
rutherfordium. Breit and quantum electrodynamic corrections are also
included. The results for the superheavy elements are compared with
earlier calculations where available. Similar calculations for lighter
analogs, ytterbium, 
lutetium, and hafnium are used to study the accuracy of the
calculations. The estimated uncertainties of the final results are
discussed.

\end{abstract}

\pacs{31.15.vj, 31.30.jg, 11.30.Er}

\maketitle

\section{Introduction}

The study of the superheavy elements (nuclear charge $Z>100$)  is an
important multidisciplinary area of research involving nuclear physics,  atomic physics, and chemistry (see,
e.g. reviews~\cite{rev1,rev2,rev3}). Atomic calculations
help to understand the role of the relativistic and many-body effects
and provide important information for the planing and interpreting
the measurements. The need to treat relativistic and correlation
effects to high level of accuracy makes the calculations a very
challenging task. Relativistic effects are most important for the
structure of the inner electron shells. Their effect on the spectra of
neutral atom, determined by valence electrons, is much
smaller. Standard approach based on using Dirac equation and adding
Breit and quantum electrodynamic (QED) corrections gives reasonably
good results (see,
e.g.~\cite{Dinh119,Dinh120,Dinh112,Dzuba119}). Accurate treatment of
correlations is a more difficult task. Most of superheavy elements
have open shells with many valence electrons and strong correlations
between them and between valence electrons and electrons in the
core. Therefore, it is particulary important to establish the benchmark
values for superheavy systems that have  one to four valence
electrons which can be treated by the most high-precision approaches.
Such calculations also establish the importance of various corrections that may be used for
more complicated superheavy atoms.
 In our previous
papers~\cite{Dinh119,Dinh120,Dinh112,Dzuba119}  we
studied the elements with nuclear charge $Z=112$, 119 and 120, which
are heavier analogs of mercury, francium and radium respectively. These systems have one or two
valence electrons.  In
present paper, we calculate the spectra and other atomic  properties of superheavy atoms with two,
 three, and four valence electrons above closed shells:
nobelium ($Z=102$), lawrencium ($Z=103$), and rutherfordium
($Z=104$). These elements are heavier analogs of ytterbium, lutetium
and hafnium. No, Lr, and Rf 
were studied  theoretically in
\cite{F-No,EK-No,LHZ-No,DF-Lr,EK-Lr,ZFF-Lr,FDKU-Lr,BEK-Lr,EK-Rf,MS-Rf} and
experimentally in \cite{rev1,rev2,rev3},  but experimental spectra  still
have not been measured. Present relativistic calculation use the combination of the configuration
interaction (CI) method with the linearized single-double coupled
cluster method (CI+SD or CI+all-order)~\cite{SD+CI}.
Correlations between valence electrons are treated with
the CI technique while correlations between core and valence electrons
are included via single-double coupled-cluster method. This approach provides the
 most complete treatment of the inter-electron
correlations since it includes core-core,  core-valence and valence-valence correlations
to all orders.
We treat nobelium, lawrencium and
rutherfordium as two-, three-, and four-valence electrons systems
respectively.
Previous calculations
for No~\cite{F-No,EK-No,LHZ-No} and Rf~\cite{EK-Rf,MS-Rf} considered
these atoms as two-valence electron systems, while calculations for
Lr~\cite{DF-Lr,EK-Lr,ZFF-Lr,FDKU-Lr,BEK-Lr}  treated the atom as a
monovalent  system. Such treatments omit important correlation effects for Lr and Rf.
 Comparing present and earlier calculations provide important
information on the role of different types of correlation and
relativistic corrections.
We also present calculations of few first ionization potentials for
No, Lr and Rf, up to removal of all valence electrons and calculate static polarizabilities for all three
atoms. 
In next section we describe the method and present results of
calculations for Yb, Lu, and Hf to illustrate the accuracy of the
method. In last section we present results and detailed discussion for
No, Lr, and Rf.

\section{Method of calculation}

The calculations are performed using the configuration
interaction method combined with the linearized single-double 
coupled-cluster method introduced in ~\cite{SD+CI}.
This CI+all-order method yielded accurate atomic properties for a number of divalent systems and trivalent Tl 
\cite{SD+CI,SafKozCla11,PorSafKoz12,SafPorCla12}.
It has been recently applied to the calculations of four-electron systems for the first time (Sn-like ions)
\cite{HCI}.

We use \textit{frozen core}  Dirac-Fock (DF) $V^{N-M}$ potential ~\cite{vnm} as the point of departure for all of our calculations, where
$N$ is the total number of electrons and $M$ is the number of valence
electrons, i.e. the initial DF procedure is
carried out  for the closed-shell ion, with all valence electrons removed. For the atoms treated here,  $M$=2 for Yb and No, $M$=3 for Lu and Lr, and  $M$=4 for Hf and
Rf.
The effective CI Hamiltonian for states of valence electrons is the sum
of single-electron Hamiltonians and an operator representing
the interaction between the valence electrons,
\begin{equation}
  \hat H^{\rm eff} = \sum_{i=1}^{M}\hat h_1(r_i) + \sum_{i<j}\hat h_2(r_i,r_j).
\label{heff}
\end{equation}

The single-electron Hamiltonian for a valence electron has the form
\begin{equation}
  \hat h_1 = c \mathbf{\alpha p} + (\beta -1)mc^2 - \frac{Ze^2}{r} +
  V^{N-M}+ \hat \Sigma_1,
\label{h0}
\end{equation}
where $\hat \Sigma_1$ is the correlation potential operator, which represents
the correlation interaction of a valence electron with the core. Its
matrix elements are related to the single-excitation amplitudes of the
coupled-cluster method via
\begin{equation}
  \Sigma_{mv} = \rho_{mv}(\widetilde{\epsilon}_{v}-\epsilon_m),
\label{sigma1}
\end{equation}
where $\rho_{mv}$ is an excitation coefficient of the atomic wave
function for the term with excitation from the valence state $v$ to
another excited state $m$; and $\epsilon_m$ are
Dirac-Fock energies of corresponding single-electron basis states.
 The quantities $\widetilde{\epsilon}_{v}$ are discussed in detail in Ref.~\cite{SD+CI}.
Briefly,  the CI+all-order approach is based on the 
Brillouin-Wigner variant of MBPT rather than the Rayleigh-Schr\"{o}dinger variant resulting in the energy dependence of the $\Sigma$.
  Ideally, the energy $\widetilde{\epsilon}_{v}$ should be
calculated from the particular eigenvalue of the effective
Hamiltonian. In actual calculations, the simplest and the most practical approach
 is to set the energy $\widetilde{\epsilon}_{v}$
 to the Dirac-Fock energy of the lowest orbital for the
particular partial wave. For example, we use $\widetilde{\epsilon}_{v}=\epsilon_{6s}$ for all
$ns$ orbitals of  Yb atom.

The interaction between valence electrons is the sum of the Coulomb interaction
and the correlation correction operator $\hat \Sigma_2$:
\begin{equation}
  \hat h_2(r_i,r_j) = \frac{e^2}{|\mathbf{r_i - r_j}|} + \hat \Sigma_2(r_i,r_j).
\label{h2}
\end{equation}
The operator $\hat \Sigma_2$ represents the screening of the Coulomb
interaction between valence electrons by core electrons.
Its matrix elements are related to the double-excitation coupled-cluster $\rho_{mnvw}$ coefficients via
\begin{equation}
  \Sigma_{mnvw} = \rho_{mnvw}(\widetilde{\epsilon}_{v}+\widetilde{\epsilon}_{w}-\epsilon_m-\epsilon_n).
\label{sigma2}
\end{equation}

The many-electron wave function for the valence electrons $\Psi$ can
be expressed as an expansion over single-determinant wave functions
\begin{equation}
  \Psi = \sum_i c_i \Phi_i(r_1,\dots,r_M).
\label{psi}
\end{equation}
The functions $\Phi_i$ are constructed from the single-electron valence basis
states calculated in the $V^{N-M}$ potential.
The coefficients $c_i$ and many-electron energies are found by
solving the matrix eigenvalue problem
\begin{equation}
  (H^{\rm eff} - E)X = 0,
\label{Schr}
\end{equation}
where $H^{\rm eff}_{ij} = \langle \Phi_i | \hat H^{\rm eff} | \Phi_j \rangle$ and
$X = \{c_1,c_2, \dots , c_n \}$.

We use the linearized coupled-cluster method to calculate the correlation correction operators
$\hat \Sigma_1$ and $\hat \Sigma_2$. The B-spline
technique~\cite{Bspline} is used to construct a
single-electron basis for calculation of $\hat \Sigma$ and for
building many-electron basis states for the CI calculations.
We use 35 $B$-splines of order
7 in a cavity of radius $R_{max} = 60 a_B$, where $a_B$ is Bohr's radius.
All sums in the all-order terms are carried out
including $l_{max}=6$ partial waves. The contributions from $l>6$ partial waves was estimated
and included into the final results.

\subsection{Breit interaction}

Breit interaction is included in present calculations using the approach
developed in Ref.~\cite{Breit1,Breit2}.
We treat Breit interaction in zero energy transfer approximation. The
Breit Hamiltonian includes magnetic interaction between moving
electrons and retardation
\begin{equation}
\hat H^{B}=-\frac{\mbox{\boldmath$\alpha$}_{1}\cdot \mbox{\boldmath$\alpha$}_{2}+
(\mbox{\boldmath$\alpha$}_{1}\cdot {\bf n})
(\mbox{\boldmath$\alpha$}_{2}\cdot {\bf n})}{2r} \ .
\label{eq:Breit}
\end{equation}
Here ${\bf r}={\bf n}r$, $r$ is the distance between electrons, and
$\mbox{\boldmath$\alpha$}$ is the Dirac matrix.

Similar to the way Coulomb interaction is used to form self-consistent
Coulomb potential, Breit interaction is used to form self-consistent
Breit potential. In other words, Breit interaction is included into
self-consistent Hartree-Fock procedure. Thus the important relaxation
effect is included. The resulting inter-electron potential in
(\ref{h0}) consist of two terms
\begin{equation}
\hat V=V^{C}+V^{B} \ ,
\end{equation}
$V^{C}$ is the Coulomb potential, $V^B$ is the Breit potential.
Coulomb interaction in the second-order correlation operator $\hat
\Sigma$ is also modified to include Breit operator (\ref{eq:Breit}).
The contribution of the
Breit interaction to the energy levels of all atoms considered here is small, 
 generally less than 100~cm$^{-1}$.
\subsection{QED corrections}

We use the radiative potential method developed in Ref.~\cite{radpot}
to include quantum radiative corrections. This potential has the form
\begin{equation}
V_{\rm rad}(r)=V_U(r)+V_g(r)+ V_e(r) \ ,
\end{equation}
where $V_U$ is the Uehling potential, $V_g$ is the potential arising from the
magnetic formfactor, and $V_e$ is the potential arising from the
electric formfactor. The $V_U$ and $V_e$ terms can be considered as
additions to nuclear potential while inclusion of $V_g$ leads to some
modification of the Dirac equation (see Ref.~\cite{radpot} for details).
We find that the QED corrections are small in comparison with the
higher-order correlation corrections and can be omitted at the present
level of accuracy. We compared the results with and without QED for No
as an illustration.

\subsection{Calculation of polarizabilities}

\label{s:pol}

Polarizabilities characterize interaction of atoms with external
electric field. The Stark energy shift of atomic state $JLn$ in the
static electric field $\varepsilon$ is given by
\begin{equation}
 \Delta E(JLn) = -\left(\alpha_0 +
   \frac{3M^2-J(J+1)}{J(2J-1)}\alpha_2\right)\frac{\varepsilon}{2}^2,
 \label{eq:Stark}
 \end{equation}
 where $\alpha_0$ and $\alpha_2$ are scalar and tensor electric-dipole
 polarizabilities, and $M$ is the projection of the total angular
 momentum $J$ on the direction of electric field.
 Scalar polarizability is given by
 \begin{equation}
 \alpha_0(JLn) =
 \frac{2}{3(2J+1)}\sum_{J^{\prime}L^{\prime}n^{\prime}} \frac{\langle
   JLn|| \mathbf{D} ||
 J^{\prime}L^{\prime}n^{\prime}\rangle^2}{E_{n^{\prime}} - E_n},
 \label{eq:alpha0}
 \end{equation}
 where $\mathbf = -e \sum_i \mathbf{r}_i$ is the electric dipole
 operator. Tensor polarizability  $\alpha_2$ is non-zero only for atomic
 states with $J \ge 1$. The expression for $\alpha_2$ differs from
 (\ref{eq:alpha0}) by an angular coefficient:
 \begin{eqnarray}
&& \alpha_2(JLn) = \sqrt{\frac{10J(2J-1)}{3(2J+3)(2J+1)(J+1)}}
\times \label{eq:alpha2}  \\
&& \sum (-1)^{(J+J^{\prime})}\left\{ \begin{array}{ccc}
 1 & 1 & 2 \\ J & J & J^{\prime} \end{array} \right\}
 \frac{\langle JLn|| \mathbf{D} ||
 J^{\prime}L^{\prime}n^{\prime}\rangle^2}{E_{n^{\prime}} - E_n},
 \nonumber
 \end{eqnarray}
 The expressions (\ref{eq:alpha0}) and (\ref{eq:alpha2}) are exact if
 $|JLn\rangle$ and $|J^{\prime}L^{\prime}n^{\prime}\rangle$ are exact
 many-electron wave functions. In practice, atomic electrons are
 divided into core and valence electrons and the expression for scalar
 polarizability  becomes a sum of three terms
 \begin{equation}
 \alpha_0 = \alpha_c + \alpha_{cv}+\alpha_{v}.
 \label{eq:alphasum}
 \end{equation}

\begin{table}
\caption{Energies ($E$, cm$^{-1}$) \cite{Yb} and $g$-factors of the lowest states of
  ytterbium. Comparison of calculations with
  experiment. Non-relativistic values of $g$-factors ($g_{\rm nr}$)
  are given by (\ref{eq:gnr}).}
\label{t:Yb}
\begin{ruledtabular}
\begin{tabular}{llrrrrrr}
\multicolumn{1}{c}{Conf.}&
\multicolumn{1}{c}{Term}&
\multicolumn{3}{c}{Energy}&
\multicolumn{3}{c}{g-factors}\\
\multicolumn{1}{c}{}&
\multicolumn{1}{c}{}&
\multicolumn{1}{c}{Expt.}&
\multicolumn{1}{c}{Present}&
\multicolumn{1}{c}{Diff.}&
\multicolumn{1}{c}{Expt.}&
\multicolumn{1}{c}{nr}&
\multicolumn{1}{c}{Present}\\
\hline
$6s^2$   & $^1S_0 $ &      0 &     0 &      0&        & 0      &  0   \\[0.4pc]

$6s6p$   &$ ^3P_{0}$&   17288&   17561&    -273&        & 0      &  0      \\
         &$ ^3P_{1}$&   17992&   18261&    -269& 1.49282& 1.5000 &  1.4921 \\
         &$ ^3P_{2}$&   19710&   20010&    -300& 1.50   & 1.5000 &  1.5000 \\[0.4pc]

$5d6s$   &$ ^3D_{1}$&   24489&   24505&     -16& 0.50   & 0.5000 &  0.5000 \\
         &$ ^3D_{2}$&   24752&   24863&    -111& 1.16   & 1.1667 &  1.1634 \\
         &$ ^3D_{3}$&   25271&   25343&     -72& 1.34   & 1.3333 &  1.3333 \\[0.4pc]

$6s6p$   & $^1P_1 $ &   25068&   25816&    -748& 1.035  & 1.0000 &  1.0087 \\[0.4pc]

$5d6s$   & $^1D_2 $ &   27678&   27991&    -313& 1.01   & 1.0000 &  1.0036 \\[0.4pc]
$6s7s$   & $^3S_1 $ &   32695&   32970&    -275& 2.01   & 2.0000 &  1.9998 \\
$6s7s$   & $^1S_0 $ &   34351&   34579&    -228&        & 0      &  0   \\[0.4pc]

$6s7p$   & $^3P_0 $ &   38091&   38377&    -286&        & 0      &  0 \\
         &$ ^3P_{1}$&   38174&   38440&    -266& 1.14   & 1.5000 &  1.4399 \\
         &$ ^3P_{2}$&   38552&   38821&    -269& 1.50   & 1.5000 &  1.4999 \\[0.4pc]

$6s6d$   &$ ^3D_{1}$&   39809&   40053&    -244& 0.50   & 0.5000 &  0.5001 \\
         &$ ^3D_{2}$&   39838&   40147&    -309& 1.16   & 1.1667 &  1.1414 \\
         &$ ^3D_{3}$&   39966&   40205&    -239& 1.33   & 1.3333 &  1.3333 \\
$6s6d$   & $^1D_2 $ &   40062&   40089&     -27& 1.03   & 1.0000 &  1.1423 \\[0.4pc]

$6s7p$   & $^1P_1 $ &   40564&   39150&    1414& 1.01   & 1.0000 &  1.0598 \\[0.4pc]
$6s8s$   & $^3S_1 $ &   41615&   41997&    -382& 2.02   & 2.0000 &  1.9994 \\
$6s8s$   & $^1S_0 $ &   41940&   42397&    -457&        & 0      &  0 \\
\end{tabular}
\footnotetext[1]{Ref. \cite{NIST}}
\end{ruledtabular}
\end{table}

\begin{table}
\caption{Energies ($E$, cm$^{-1}$) and $g$-factors of the lowest states of
  lutetium. Comparison of calculations with
  experiment. Non-relativistic values of $g$-factors ($g_{\rm nr}$)
  are given by (\ref{eq:gnr}).}
\label{t:Lu}
\begin{ruledtabular}
\begin{tabular}{llrrrrrr}
\multicolumn{1}{c}{Conf.}&
\multicolumn{1}{c}{Term}&
\multicolumn{3}{c}{Energy}&
\multicolumn{3}{c}{g-factors}\\
\multicolumn{1}{c}{}&
\multicolumn{1}{c}{}&
\multicolumn{1}{c}{Expt.}&
\multicolumn{1}{c}{Present}&
\multicolumn{1}{c}{Diff.}&
\multicolumn{1}{c}{Expt.}&
\multicolumn{1}{c}{nr}&
\multicolumn{1}{c}{Present}\\
\hline
$5d6s^2$  & $^2D_{3/2}$&      0 &       0&       0& 0.79921& 0.8000 &  0.8000 \\
          & $^2D_{5/2}$&    1993&   2014 &    -21&  1.20040& 1.2000 &  1.1999 \\[0.4pc]

$6s^26p$  & $^2P_{1/2}$&    4136&   3910 &    226&  0.66   & 0.6666 &  0.6661 \\
          & $^2P_{3/2}$&    7476&   7228 &    248&  1.33   & 1.3333 &  1.3333 \\[0.4pc]

$5d6s6p$  & $^4F_{3/2}$&   17427&  17723 &   -296&  0.59   & 0.4000 &  0.4525 \\
          & $^4F_{5/2}$&   18504&  18789 &   -285&  1.07   & 1.0286 &  1.0586 \\
          & $^4F_{7/2}$&   20432&  20731 &   -299&  1.22   & 1.2381 &  1.2424 \\
          & $^4F_{9/2}$&   22609&  22911 &   -302&  1.30   & 1.3333 &  1.3332 \\[0.4pc]

$5d^26s$  & $^4F_{3/2}$&   18851&  19182 &   -331&         & 0.4000 &  0.4109 \\
          & $^4F_{5/2}$&   19403&  19737 &   -334&         & 1.0286 &   1.0305 \\
          & $^4F_{7/2}$&   20247&  20578 &   -331&         & 1.2381 &   1.2368 \\
          & $^4F_{9/2}$&   21242&  21591 &   -349&  1.0    & 1.3333 &   1.3313 \\[0.4pc]

$5d6s6p$  & $^4D_{1/2}$&   20762&  20995 &   -233&  0.00   & 0.0000 &  0.0353 \\
          & $^4D_{3/2}$&   21195&  21448 &   -253&  1.19   & 1.2000 &  1.1551 \\
          & $^4D_{5/2}$&   22221&  22504 &   -283&  1.39   & 1.3714 &  1.3799 \\
          & $^4D_{7/2}$&   23524&  23795 &   -271&  1.41   & 1.4286 &  1.4171 \\[0.4pc]

$5d6s6p$  & $^2D_{5/2}$&   21462&  21735 &   -273&  1.23   & 1.2000 &  1.2107  \\
          & $^2D_{3/2}$&   22124&  22376 &   -252&  0.874  & 0.8000 &  0.8591 \\[0.4pc]

$5d^26s$  & $^4P_{1/2}$&   21472&  21860 &   -388&         & 2.6667 &  2.6098 \\
          & $^4P_{3/2}$&   22467&  22849 &   -382&  1.73   & 1.7333 &  1.7016 \\
          & $^4P_{5/2}$&   22802&  23242 &   -440&         & 1.6000 &  1.4749 \\[0.4pc]

$5d6s6p$  & $^4P_{1/2}$&   24108&  24520 &   -412&         & 2.6667 &  2.6264 \\
          & $^4P_{3/2}$&   24308&  24786 &   -478&  1.67   & 1.7333 &  1.6530 \\
          & $^4P_{5/2}$&   25191&  25774 &   -583&  1.53   & 1.6000 &  1.5267 \\[0.4pc]

$5d^26s$  & $^2D_{3/2}$&   24518&  25015 &   -497&         & 0.8000 &  0.8379 \\
\end{tabular}
\footnotetext[1]{Ref. \cite{NIST}}
\end{ruledtabular}
\end{table}

 \begin{table}
\caption{\label{t:Hf}
 Energies ($E$, cm$^{-1}$) and
$g$-factors of the lowest states of
  hafnium. Non-relativistic values of $g$-factors ($g_{\rm nr}$)
  are given by Eq.(\ref{eq:gnr})
    Comparison of calculations with experiment \cite{NIST}.
  Result with ``*'' is by  Sansonetti and  Martin \cite{Martin}.
 }
\begin{ruledtabular}
\begin{tabular}{llrrrrrr}
\multicolumn{1}{c}{Conf.}&
\multicolumn{1}{c}{Term}&
\multicolumn{3}{c}{Energy}&
\multicolumn{3}{c}{g-factors}\\
\multicolumn{1}{c}{}&
\multicolumn{1}{c}{}&
\multicolumn{1}{c}{Expt.}&
\multicolumn{1}{c}{Present}&
\multicolumn{1}{c}{Diff.}&
\multicolumn{1}{c}{Expt.}&
\multicolumn{1}{c}{nr}&
\multicolumn{1}{c}{Present}\\
\hline
 $5d^26s^2$&$  ^3F_{2}$&       0.&        0&         &   0.695 &  0.667 &  0.6936  \\
           &$  ^3F_{3}$&    2357 &     2343&      14 &   1.083 &  1.083 &  1.0832  \\
           &$  ^3F_{4}$&    4568 &     4617&     -49 &   1.240 &  1.250 &  1.2425  \\[0.4pc]

 $5d^26s^2$&$  ^3P_{0}$&    5522 &     5611&     -89 &   0.00   &  0.00   &  0.00    \\
           &$  ^3P_{1}$&    6573 &     6594&     -21 &   1.500 &  1.500 &  1.5000  \\
           &$  ^3P_{2}$&    8984 &     9151&    -167 &   1.300 &  1.500 &  1.2783  \\[0.4pc]

 $5d^26s^2$&$  ^1D_{2}$&    5639&      5842&    -203 &    1.165 &  1.000 &  1.1947  \\[0.4pc]

 $5d6s^26p$&$  ^1D_{2}$&   10509$^*$& 10095&     414 &          &   1.000 &  0.8173 \\[0.4pc]

 $5d^26s^2$&$  ^1G_{4}$&   10533 &    11411&    -878 &   1.008 &  1.000 &  1.0073  \\[0.4pc]

 $5d6s^26p$&$  ^3D_{1}$&   14018&     13718&     300 &    0.55  &  0.500  &  0.5384 \\
           &$  ^3D_{2}$&   16163&     15840&     323 &    1.17  &  1.167  &  1.1714 \\
           &$  ^3D_{3}$&   18381&     18084&     297 &    1.29  &  1.333  &  1.2980 \\[0.4pc]

 $5d^36s  $&$  ^5F_{1}$&   14092 &    14445&    -353 &   0.00  &  0.00   &   0.0217  \\
           &$  ^5F_{2}$&   14741 &    15079&    -338 &   1.00  &  1.000  &  1.0038  \\
           &$  ^5F_{3}$&   15673 &    15996&    -323 &   1.25  &  1.250  &  1.2485  \\
           &$  ^5F_{4}$&   16767 &    17099&    -332 &   1.36  &  1.350  &  1.3445  \\[0.4pc]

 $5d6s^26p$&$  ^3F_{2}$&   14435&     14019&     416 &    0.89  &  0.666  &  0.8914 \\
           &$  ^3F_{3}$&   14542&     14210&     332 &    1.08  &  1.083  &  1.0877 \\
           &$  ^3F_{4}$&   18225&     17887&     338 &    1.24  &  1.250  &  1.2451 \\[0.4pc]

 $5d6s^26p$&$  ^3P_{1}$&   18143&     17932&     211 &    1.43  &  1.500  &  1.4401 \\
           &$  ^3P_{2}$&   19791&     19584&     207 &    1.41  &  1.500  &  1.4192 \\[0.4pc]

 $5d^26s6p$&$  ^5G_{2}$&   18011 &    17996&      15 &   0.40  &  0.333   &  0.3874 \\
           &$  ^5G_{3}$&   19293 &    19262&      31 &   0.95  &  0.917   &  0.9375 \\
           &$  ^5G_{4}$&   20960 &    20935&      25 &   1.16  &  1.150   &  1.1597 \\
\end{tabular}
\end{ruledtabular}
\end{table}

 \begin{table}
\caption{\label{t:No} Energies ($E$, cm$^{-1}$) and $g$-factors of
the lowest states of Nobelium.  Non-relativistic values of
$g$-factors ($g_{\rm nr}$)
  are given by Eq.(\ref{eq:gnr}). Comparison with theoretical results
presented by Borschevsky  {\it et al.\/} ~\cite{EK-No} and Liu
{\it et al.\/} \cite{LHZ-No}. }
\begin{ruledtabular}
\begin{tabular}{lrrrrrrr}
\multicolumn{1}{c}{Conf.}&
\multicolumn{1}{c}{Term}&
\multicolumn{2}{c}{Energy}&
\multicolumn{2}{c}{Energy}&
\multicolumn{2}{c}{g-factors}\\
\multicolumn{1}{c}{}&
\multicolumn{1}{c}{}&
\multicolumn{1}{c}{Present}&
\multicolumn{1}{c}{+Lamb}&
\multicolumn{1}{c}{Ref.~\cite{EK-No}} &
\multicolumn{1}{c}{Ref.~\cite{LHZ-No}}&
\multicolumn{1}{c}{Present}&
\multicolumn{1}{c}{nr}\\
\hline
$7s2 $&$    ^1S_{0}$&         0&          0&        0&        0&        0&      0   \\[0.4pc]
$7s7p$&$    ^3P_{0}$&     19682&      19567&    18879&    19798&        0 &     0    \\
      &$    ^3P_{1}$&     21156&      21042&    20454&    21329&   1.4577&    1.500 \\
      &$    ^3P_{2}$&     26225&      26113&    25374&    26186&   1.4998&    1.500 \\[0.4pc]

$7s7p$&$    ^1P_{1}$&     30304&      30203&    30056&    30069&   1.0409&    1.000 \\[0.4pc]

$7s6d$&$    ^3D_{1}$&     28587&      28436&    28338&&   0.5000&    0.500 \\
      &$    ^3D_{2}$&     29098&      28942&    28778&&   1.1606&    1.167 \\
      &$    ^3D_{3}$&     30322&      30183&    29897&&   1.3332&    1.333 \\[0.4pc]

$7s6d$&$    ^1D_{2}$&     33657&      33504&    32892&&   1.0071&    1.000 \\[0.4pc]

$7s8s$&$    ^3S_{1}$&     35815&      35731&    35092&&   1.9994&    2.000 \\[0.4pc]

$7s8s$&$    ^1S_{0}$&     37444&      37360&    36538&&   0.0000&    0.000   \\[0.4pc]

$7s8p$&$    ^3P_{0}$&    41365&      41266&    40576&&   0.0000&    0.000   \\
      &$    ^3P_{1}$&    41481&      41382&    40692&&   1.4083&    1.500 \\
      &$    ^3P_{2}$&    42582&      42484&    42837&&   1.4999&    1.500 \\[0.4pc]

$7s8p$&$    ^1P_{1}$&    43011&      42910&    42285&&   1.0917&    1.000 \\

$7s7d$&$    ^3D_{1}$&    43522&      43422&    42726&&   0.5002&    0.500 \\
      &$    ^3D_{2}$&    43581&      43481&    42758&&   1.1452&    1.167 \\
      &$    ^3D_{3}$&    43830&      43730&    43033&&   1.3333&    1.333 \\[0.4pc]

$7s7d$&$    ^1D_{2}$&    44099&      43999&    43079&&   1.0216&    1.000 \\[0.4pc]
$7s9s$&$    ^3S_{1}$&    44894&      44794&    44247&&   1.9994&    2.000 \\[0.4pc]

$7s6f$&$    ^3F_{2}$&    46795&      46695&    &      &   0.6669&    0.667\\
      &$    ^3F_{3}$&    46788&      46688&    &      &   1.0072&    1.083 \\
      &$    ^3F_{4}$&    46810&      46710&    &      &   1.2500&    1.250 \\[0.4pc]

$7s6f$&$    ^1F_{3}$&    46806&      46706&    &      &   1.0762&    1.000 \\

 \end{tabular}
\end{ruledtabular}
\end{table}


\begin{table*}
\caption{Calculated energies ($E$, cm$^{-1}$) and $g$-factors of the
  lowest states of lawrencium. Comparison with other calculations.
  Non-relativistic values of $g$-factors ($g_{\rm nr}$)
  are given by (\ref{eq:gnr}).}
\label{t:Lr}
\begin{ruledtabular}
\begin{tabular}{llc rl lrrrr}
Config. & Term & $J$ & \multicolumn{3}{c}{Present work} &
 \multicolumn{4}{c}{Other energy} \\
&&&\multicolumn{1}{c}{Energy} &
\multicolumn{1}{c}{$g$} &
\multicolumn{1}{c}{$g_{\rm nr}$} &
\multicolumn{1}{c}{Ref.~\cite{EK-Lr}} &
\multicolumn{1}{c}{Ref.~\cite{ZFF-Lr}} &
\multicolumn{1}{c}{Ref.~\cite{FDKU-Lr}} &
\multicolumn{1}{c}{Ref.~\cite{BEK-Lr}} \\
\hline
$7s^27p$ & $^2$P$^o$ & 1/2 &    0 & 0.6652 & 0.6666 & 0 & 0 & 0& 0\\
         &           & 3/2 & 8495 & 1.3333 & 1.3333 & 8273 & 8935 & 8138 & 8389 \\[0.4pc]

$7s^26d$ & $^2$D    & 3/2 &  1555 & 0.8002 & 0.8000 & 1263 & 1127 & 1331 & 1408 \\
         &          & 5/2 &  5423 & 1.2001 & 1.2000 & 5062 & & 4187 & 5082 \\[0.4pc]

$7s7p6d$ & $^4$F$^o$ & 3/2 & 21288 & 0.4803 & 0.4000 & & & 20886 &  \\
         &          & 5/2 & 23530 & 1.0668 & 1.0286 &  & & 23155 & \\
         &          & 7/2 &       &        & 1.2381 &  & & 27276 & \\
         &          & 9/2 &       &        & 1.3333 &  & & 32775 & \\[0.4pc]

$7s^28s$ & $^2$S    & 1/2 & 20253 & 2.0163 & 2.0000 & & & 20405 & 20131 \\[0.4pc]

$7s^28p$ & $^2$P$^o$ & 1/2 & 25912& 0.6161 & 0.6666 & & & & 26104\\
          &          & 3/2 & 27079 & 1.3174 & 1.3333 & & & & 27491\\[0.4pc]

$7s6d^2$ & $^4$P    & 1/2 & 25409 & 2.4737 & 2.6667 & & & & \\
$7s6d^2$ &          & 3/2 & 26327 & 1.5286 & 1.7333 & & & & \\
$7s6d^2$ &          & 5/2 & 27397 & 1.3148 & 1.6000 & & & & \\





\end{tabular}
\end{ruledtabular}
\end{table*}


 \begin{table}
\caption{\label{t:Rf}  Energies ($E$, cm$^{-1}$) and
$g$-factors of the lowest states of Rutherfordium.
  Non-relativistic values of $g$-factors ($g_{\rm nr}$)
  are given by (\ref{eq:gnr}).
Comparison with results by Kaldor \cite{EK-Rf}.
 }
\begin{ruledtabular}
\begin{tabular}{lrrrrrrr}
\multicolumn{1}{c}{Conf.}&
\multicolumn{1}{c}{Term}&
\multicolumn{2}{c}{Energy}&
\multicolumn{2}{c}{g-factors}\\
\multicolumn{1}{c}{}&
\multicolumn{1}{c}{}&
\multicolumn{1}{c}{Present}&
\multicolumn{1}{c}{ \cite{EK-Rf}}&
\multicolumn{1}{c}{Present}&
\multicolumn{1}{c}{nr}\\
\hline
 $ 7s^26d^2$&$   ^3F_{2} $&       0&     0    & 0.7291& 0.667\\
            &$   ^3F_{3} $&    4904&    4855  & 1.0834& 1.083 \\
            &$   ^3F_{4} $&    8625&    7542  & 1.2062&  1.250 \\[0.4pc]

 $ 7s^27p6d$&$   ^3F_{2} $&    2547&     2210 &  0.7869&  0.667 \\
            &$   ^3F_{3} $&   11390&    11905 &  1.1041&  1.083 \\
            &$    3F_{4} $&   20477&          &  1.2489&  1.250 \\[0.4pc]

 $ 7s^26d^2$&$   ^3P_{0} $&    5034&          &     0.0&    0.0    \\
            &$   ^3P_{1} $&    8348&    8776  &  1.4996&  1.500 \\
            &$   ^3P_{2} $&    7398&    7542  &  1.1853&  1.500 \\[0.4pc]

 $ 7s^27p6d$&$   ^3D_{1} $&    8288&     8373 &  0.6794&  0.500 \\
           &$   ^3D_{2} $&   11273&    10905 &  1.1493&  1.167 \\
           &$   ^3D_{3} $&   18029&          &  1.2016&  1.333 \\[0.4pc]

 $ 7s^26d^2$&$   ^1D_{2} $&   13630&          &  1.2531&  1.000 \\[0.4pc]

 $ 7s^26d^2$&$   ^1G_{4} $&   14476&          &  1.0439&  1.000 \\[0.4pc]

 $ 7s^27p6d$&$   ^1D_{2} $&   14403&          &  1.0650&  1.000 \\[0.4pc]

 $ 7s^26d^2$&$   ^1S_{0} $&   18679&          &  0.0   &  0.0       \\[0.4pc]

 $ 7s^27p6d$&$   ^1F_{3} $&   24634&          &  1.1077&   1.000      \\[0.4pc]

 $ 7s6d^3  $&$   ^5F_{1} $&   21552&          &  0.0962&  0.000 \\
            &$   ^5F_{2} $&   23079&          &  1.0289&  1.000 \\
            &$   ^5F_{3} $&   25432&          &  1.2475&  1.250 \\[0.4pc]

 $ 7s^27p6d$&$   ^3P_{1} $&   16551&          &  1.2712&  1.500 \\
            &$   ^3P_{2} $&   21480&          &  1.2267&  1.500 \\[0.4pc]

 $ 7s6d^27p$&$   ^5G_{2} $&   20347&          &  0.5067&  0.333\\
            &$   ^5G_{3} $&   23325&          &  0.9523&  0.917 \\
\end{tabular}
\end{ruledtabular}
\end{table}

Here $\alpha_c$ is the polarizability of atomic core,  $\alpha_{cv}$
is the contribution caused by Pauli principle which implies that the
excitations from the core cannot go into occupied valence
states. Therefore, polarizability of the core is different for the
ion, which has no valence electrons and for the neutral atom. This
difference is separated into $\alpha_{cv}$. Usually this contribution
is small and needs to be taken into account only in very precise
calculations. We neglect it in present work. The term $\alpha_v$
is the dominant contribution due to the valence electrons.  The core
contribution is given by
\begin{equation}
 \alpha_c = \frac{2}{3}\sum_{cm} \frac{\langle cm || \mathbf{d} || m\rangle
 \langle m || \mathbf{d} +\mathbf{\delta V}_{core} ||
 c\rangle}{\epsilon_c - \epsilon_m},
 \label{eq:alphac}
 \end{equation}
where summation goes over core states $c$ and a complete set of
single-electron states $m$. The energies $\epsilon_c$ and $\epsilon_m$
are the single-electron energies of the basis states. The operator
$\mathbf{d}=-e\mathbf{r}$ in (\ref{eq:alphac}) is the single-electron
electric dipole operator, $\mathbf{\delta V}_{core}$ is the correction
to the self-consistent
core potential due to the effect of electric field. It is also known
as the core polarization correction or random-phase approximation
(RPA) correction. This correction is calculated  by
solving the RPA-type equations for atomic core
\begin{equation}
(\hat H_0 - \epsilon_c)\delta \psi_c = - \psi_c (\mathbf{d} +
\mathbf{\delta V}_{core}),
\label{eq:RPA}
\end{equation}
where $\hat H_0$ is the Hartree-Fock Hamiltonian, $\delta \psi_c$ is
the correction to the core state $\psi_c$ due to the effect of
external electric field. The equations (\ref{eq:RPA}) are solved
self-consistently for all states in the core and the correction to the
core potential $\delta V_{core}$ is found. Core contribution is small, ranging from 3.20~a.u. for Hf
 to  $8.46$~a.u. for No. 
 The
core does not contribute to the tensor polarizability since the
total angular momentum of the closed shell core is zero.

The expressions for the  valence contributions to the scalar
and tensor polarizabilities are very similar to  (\ref{eq:alpha0})
and (\ref{eq:alpha2}) with a few modifications. The many-electron
states  $|JLn\rangle$ and $|J^{\prime}L^{\prime}n^{\prime}\rangle$ are
now the valence states, the summation in the
electric dipole operator $\mathbf{D}$ goes over only valence electrons,
 and every  single-electron electric dipole operator $d$ is
modified to include core polarization correction,
$\mathbf{\tilde d} = \mathbf{d} + \mathbf{\delta V}_{core}$.

To perform summation in (\ref{eq:alpha0}) and (\ref{eq:alpha2}) over
complete set of many-valence-electrons  states we use method suggested
by Dargarno and Lewis~\cite{DalLew55}. The summation is reduced to
calculation of the correction to the ground state wave function
\begin{equation}
\sum_n \frac{\langle a || \tilde D || n\rangle \langle n ||\tilde D||
  a|\rangle}{E_a-E_n} = \langle a || \tilde D || \tilde a \rangle,
\label{eq:Dal}
\end{equation}
where correction $|\tilde a\rangle$ to the ground state wave function
$|a\rangle$ is given by
\begin{equation}
|\tilde a\rangle = \sum_n |b\rangle\frac{\langle a || \tilde D ||
  n\rangle }{E_a-E_n}.
\label{eq:Dal1}
\end{equation}
The correction  $|\tilde a\rangle$ satisfies the inhomogeneous equation
\begin{equation}
(H^{\rm CI} - E_a) |\tilde a \rangle = - \tilde D H^{\rm CI}.
\label{eq:da}
\end{equation}
Here $H^{\rm CI}$ is the effective CI Hamiltonian presented in a
matrix form while $|\tilde a\rangle$ is a vector of expansion
coefficients over single-determinant basis states. Solving the system
of linear equations (\ref{eq:da}) and substituting the result into
(\ref{eq:Dal}) is equivalent to summation over all possible many
electron states which can be constructed from the given
single-electron basis.

\subsection{Results for Yb, Lu and Hf}

\begin{table}
\caption{\label{tab-corr}  Comparison of higher-order (III$^+$) correlation contributions to three-electron removal energies of
Lu and Lr. Columns CI+MBPT and CI+sll give removal energies calculated in the
respective approximations. The difference of these values give the estimate of the
 higher-order correlation correction given in column labeled Diff. Last column labelled Diff$^*$ give the
 higher-order correlation relative to the corresponding ground states.}
\begin{ruledtabular}
\begin{tabular}{llcccr}
   \multicolumn{1}{c}{Atom}&
   \multicolumn{1}{l}{Level}&
\multicolumn{1}{c}{CI+MBPT}&
\multicolumn{1}{c}{CI+all}&
\multicolumn{1}{c}{Diff}&
  \multicolumn{1}{c}{Diff$^*$}\\
  \hline
Lu& $6s^2 5d~^2D_{3/2}$ &  328791 & 325983 & -2808 &  0      \\
  & $6s^2 5d~^2D_{5/2}$ &  326610 & 323981 & -2629 &  179    \\
  & $5d^2 6s~^4F_{3/2}$ &  309931 & 306716 & -3215 &  -406   \\
  & $5d^2 6s~^4F_{5/2}$ &  309307 & 306166 & -3141 &  -333   \\
  & $5d^2 6s~^4F_{7/2}$ &  308356 & 305330 & -3026 &  -217   \\
  & $5d^2 6s~^4F_{9/2}$ &  307222 & 304323 & -2898 &  -90    \\ [0.3pc]
  & $6s^2 6p~ ^2P_{1/2}$&  324182 & 322187 & -1996 &  812    \\
  & $6s^2 6p~ ^2P_{3/2}$&  320859 & 318866 & -1993 &  815    \\
  & $5d6s6p ~ ^4F_{3/2}$&  310376 & 308268 & -2107 &  701    \\ [0.5pc]
Lr&$7s^2 6d~  ^2D_{3/2}$&  337828&   331718&   -6110&  -1422  \\
  &$7s^2 6d~  ^2D_{5/2}$&  333287&   327872&   -5415&  -726   \\
  &$7s^2 8s~  ^2S_{1/2}$&  318449&   313137&   -5311&  -623   \\
  &$7s6d^2 ~  ^4P_{1/2}$&  311742&   307921&   -3821&  867    \\
  &$7s6d^2 ~  ^4P_{3/2}$&  313159&   306879&   -6280&  -1592  \\
  &$7s6d^2 ~  ^4P_{5/2}$&  311830&   305825&   -6005&  -1317  \\  [0.3pc]
  &$7s^2 7p~  ^2P_{1/2}$&  338055&   333366&   -4688&  0     \\
  &$7s^2 7p~  ^2P_{3/2}$&  329645&   324877&   -4768&  -80   \\
  &$7s7p6d ~  ^4F_{3/2}$&  316309&   311992&   -4318&  371   \\
  &$7s7p6d ~  ^4F_{5/2}$&  314099&   309752&   -4348&  341   \\
  &$7s^28p ~  ^2P_{1/2}$&  312532&   307454&   -5079&  -391  \\
  &$7s^28p ~  ^2p_{3/2}$&  311250&   306266&   -4984&  -296  \\
     \end{tabular}
\end{ruledtabular}
\end{table}

First, we carry out the calculations for ``reference'' atoms Yb, Lu and Hf
which have the closest electronic structure to the superheavy
No, Lr, and Rf, respectively. Since the energies of Yb, Lu, and Hf
are known experimentally, such calculation provides the benchmark test of our method. Further comparison of
correlation corrections in ``reference'' and superheavy atoms allow to predict the accuracy of our
approach for superheavy elements.

Our calculated  energy levels of Yb, Lu and Hf are
compared with experiment  in Tables \ref{t:Yb}, \ref{t:Lu}, and \ref{t:Hf}, respectively.
Yb energy levels have been presented in Ref.~\cite{Yb}. Comparison  shows
 that relative theoretical
error in excitation energies is rather similar for Yb, Lu, and Hf, with somewhat
better accuracy for two-electron Yb.

The difference with experiment is $0.1-1.6$\% for Yb with the exception of the singlet
$6s6p~^1P_1$ and $6s7p~^1P_1$ states, where it is 3-3.5\%.   The lower accuracy of the singlet states arises when there is significant
  difference in the position of the triplet and singlet energy levels, such as in $^1P_1$ and $^3P_1$ case.
 It is $1.1-2.2$\% for Lu, with the exception of
  $6s^2 6p$ states where it is $3-5$\%.

 A common problem of the CI calculation with four valence electrons  is
 rapid increase of the number
of many-electron basis states with the increase in the number of
valence electrons usually leading to omitting configurations which correspond to multiple excitations of
valence electrons from the ground state to high-lying states. This
helps to reduce the CI matrix to a manageable size but leads to deterioration of the
accuracy of the calculations. However, we were able to saturate the
four-electron CI space by carrying out several very large CI calculations with diffident types of excitations,
then selecting the configuration with the largest weights from each of the runs, and combining them to produce
nearly complete configuration space. Comparing results with increasing number of selected important configurations
produced estimated uncertainty due to configuration space of less than 50~cm$^{-1}$ for most states.
 As a result, we do not observe significant deterioration of results between Lu and Hf.
 The difference with experiment is $0.1-2.9$\% for Hf with the exception of the singlet $^1D_2$ and $^1G_4$ states,
  where it is 4\% and 8\%, respectively.

We also present the values of
calculated and non-relativistic magnetic $g$-factors in Tables~\ref{t:Yb}, \ref{t:Lu}, \ref{t:Hf}.
Non-relativistic (nr) values are given by

\begin{equation}
  g_{\rm nr} = 1 + \frac{J(J+1)-L(L+1)+S(S+1)}{2J(J+1)},
\label{eq:gnr}
\end{equation}
where $J$ is total angular momentum of the atom, $L$ is its angular
momentum and $S$ is the spin ($\mathbf{J} = \mathbf{L}
+\mathbf{S}$). The $g$-factors are useful for identification of the
states.

\section{Results and discussion}

\subsection{Energy levels of No, Lr and Rf and estimates of their uncertainties}

Calculated energy levels and magnetic $g$-factors for No, Lr and Rf
are presented in Tables \ref{t:No}, \ref{t:Lr} and \ref{t:Rf} together
with the results of earlier
calculations
\cite{F-No,EK-No,LHZ-No,DF-Lr,EK-Lr,ZFF-Lr,FDKU-Lr,BEK-Lr,EK-Rf,MS-Rf}. 
We observe good agreement between the theoretical results for most of
the states. We compare No results with and without the QED correction
in Table~\ref{t:No}. The value of the QED correction is of the order
of 100 cm$^{-1}$ for most of the states while maximum value of the
correction is slightly larger than 200 cm$^{-1}$. This is smaller than
the uncertanty due to correlations (see the difference between theory
and experiment for Yb, Lu and Hf in Tables~\ref{t:Yb}, \ref{t:Lu},
\ref{t:Hf}). Therfore, we do not include QED corrections for Lr and Rf.

The accuracy of the calculations was discussed in previous section for
the case of Yb, Lu and Hf atoms. However, since relativistic and
correlations effects are larger in the superheavy elements it is
reasonable to assume that the uncertainty for No, Lr and Rf are
slightly larger than for Yb, Lu and Hf. We verified that the
contribution of the Breit interaction to the energy levels in Tables
\ref{t:Yb} - \ref{t:Rf} is small in all cases  (generally less than
100~cm$^{-1}$).  
 
  To estimate the accuracy of our values, we
directly compare the correlation effects
in Lu and Lr, since these dominate the uncertainty of the
calculations. We carry out an additional calculations for 
both atoms using a combination of the configuration interaction and second-order
many-body perturbation theory (CI+MBPT) methods ~\cite{DzuFlaKoz96b}.
In this approach, the $\Sigma_1$ and $\Sigma_2$ of the effective
Hamiltonian are build using the 
second-order perturbation theory instead of the coupled-cluster method.
The difference of the CI+MBPT and CI+all-order values gives the
estimate of the third and higher-order corrections. 
We note that Lu and Lr have different types of the ground state
configuration, $6s^2 5d$~$^2D_{3/2}$ and $7s^2
7p$~$^2P_{1/2}$. Therefore, we first directly compare the higher-order
correlation contributions to the three-electron removal energies of Lu
and Lr which are given in Table~\ref{tab-corr}. Columns CI+MBPT and
CI+all give trivalent removal energies calculated in the 
respective approximations. The difference of these values give the
estimate of the higher-order correlation correction given in column
labeled ``Diff''. 
  Last column labelled ``Diff$^*$'' gives the
 higher-order correlations relative to the corresponding ground states.
 We find that while the energies are similar for Lu and Lr, the
 higher-order correlation corrections significantly increases from  
 Lu (2000 - 3200~cm$^{-1}$)  to Lr (3800 - 6300~cm$^{-1}$). However,
 we observe that the correlation increases for all of the states and  
 when the ground state values are subtracted out, the remaining
 higher-order corrections, listed in the last column of
 Table~\ref{tab-corr} 
 are very similar for Lu and Lr. Only for the three states, $7s^26d
 ~^2D_{3/2}$ and $7s6d^2$~$^4D_{3/2, 5/2}$, the remaining
 contributions are larger than for Lu cases, which may result in
 somewhat lower accuracy for there states. 
 
Therefore, we expect $1-2$\% accuracy of the energy levels in No,
$1-3$\% in Lr, and $2-5$\% for Rf for  most of the states presented
here.



\subsection{Ionization potentials}

\label{s:IP}

Calculations in the $V^{N-M}$ approximation are very similar for a
neutral atom, negative and positive ions~\cite{vnm}. The number of
valence electrons is the only parameter in the effective CI
Hamiltonian (\ref{heff}) which changes while moving from a neutral
atom to an ion or from one ion to another. All other terms, including
Coulomb potential created by core electrons and correlation operator
$\hat \Sigma$ remain the same.
Removing one
electron from a neutral atom and comparing the energy of resulting
ground state with the energy of the ground state of neutral atom give
first ionization potential of the atom. Removing one more electron
leads to second ionization potential, etc. This process can be repeated
until all valence electrons are removed. The number of ionization
potentials which can be calculated this way is limited by the number
of valence electrons. To illustrate the accuracy of the calculations
we calculate ionization potentials for Yb, Lu
and Hf and compare them with experiment. The results are presented in
Table~\ref{t:iop}. Then in the same table we present ionization potentials
for No, Lr and Rf.

\begin{table}
\caption{Calculated ground state energies ($E_M$) of Yb, Lu, Hf, No,
  Lr and Rf neutral atoms and positive ions. $M$ is the number of
  valence electrons. The
  difference $\Delta E = E_{M-1}-E_M$ gives the ionization potential.}
\label{t:iop}
\begin{ruledtabular}
\begin{tabular}{lll ccrr}

Atom & Configu-& Term & $M$ & \multicolumn{1}{c}{$E_{M}$} &
\multicolumn{1}{c}{$\Delta E$} & Expt.\footnotemark[1] \\
/Ion & ration  &      &     & \multicolumn{1}{c}{[a.u.]}    &
\multicolumn{1}{c}{[cm$^{-1}$]}  & \multicolumn{1}{c}{[cm$^{-1}$]} \\

\hline

Yb~I  & $6s^2$   & $^1$S$_{0}$   & 2 &-0.68232  & 50768 & 50443 \\
Yb~II & $6s$   & $^2$S$_{1/2}$   & 1 &-0.45101 & 98985 & 98207 \\

Lu~I  & $6s^25d$   & $^2$D$_{3/2}$   & 3 &-1.48938 & 43289 & 43763 \\
Lu~II & $6s^2$   & $^1$S$_{0}$   & 2 &-1.29215 & 113323 & 112000 \\
Lu~III & $6s$   & $^2$S$_{1/2}$    & 1 &-0.77581 & 170270 & 169014 \\

Hf~I  & $6s^25d^2$   & $^3$F$_{0}$   & 4 & -2.83907 & 53431 & 55048 \\
Hf~II & $6s^25d$   & $^2$D$_{3/2}$   & 3 &  -2.59562 & 126748 & 120000 \\
Hf~III & $6s^2$   & $^1$S$_{0}$   & 2 &  -2.01811 & 190885 & 187800 \\
Hf~IV & $6s$   & $^2$S$_{1/2}$      & 1 &  -1.14837 & 252037 & 269150 \\

No~I  & $7s^2$   & $^1$S$_{0}$   & 2 &-0.72799   & 54390 &  \\
No~II & $7s$   & $^2$S$_{1/2}$   & 1 &-0.48018 & 105387 &  \\

Lr~I  & $7s^27p$   & $^2$P$^o_{1/2}$   & 3 &-1.52543 & 39801 &  \\
Lr~II & $7s^2$   & $^1$S$_{0}$   & 2 &-1.34408 & 118058 &  \\
Lr~III & $7s$   & $^2$S$_{1/2}$    & 1 &-0.80617 & 176934 &  \\

Rf~I  & $7s^26d^2$   & $^3$F$_{0}$   & 4 & -2.79968 & 46067 &  \\
Rf~II & $7s^25d$   & $^2$D$_{3/2}$   & 3 &  -2.58979 & 116925 &  \\
Rf~III & $7s^2$   & $^1$S$_{0}$   & 2 &  -2.05704 & 193246 &  \\
Rf~IV & $7s$   & $^2$S$_{1/2}$      & 1 &  -1.17654 & 258220 &  \\

\end{tabular}
\footnotetext[1]{Ref.~\cite{NIST}.}
\end{ruledtabular}
\end{table}

\subsection{Static polarizabilities}

\label{s:PolRes}

Results of calculations of static polarizabilities of Yb, Lu, Hf, No,
Lr and Rf are presented in Table \ref{t:pol}. 
CI+MBPT and CI+all-order results are listed in columns labelled ``MBPT'' and ``All-order'', respectively. The calculations are
done as described in section \ref{s:pol}. The result for ytterbium
agrees precisely with our previous calculations \cite{DD10,DKF14,Yb},
with experimental constrain presented in Ref.~\cite{Beloy}, and with
most of other accurate calculations (see, e. g. review~\cite{Mitroy}),
the results for lutetium and hafnium agree well with the calculations
of Doolen ~\cite{Miller}.  Estimation of accuracy is based on
comparison of the results obtained with the use of different
approaches, including comparison with experiment for ytterbium, and on
the sensitivity of the results to variation of the parameters of the
computational procedure. The theoretical uncertainty presented in the
parentheses is on the level of 5\% for Yb, Lu and Hf (see
Table~\ref{t:pol}). We expect similar uncertainty for No and
Rf. Lawrencium represents a special case due to anomalously small
energy interval between ground  $^2$P$_{1/2}$ state and first excited
$^2$D$_{3/2}$ state. Note that there is an inversion of the order of
these states in Lr as compared to its lighter analog Lu. The inversion
is due to relativistic
effects~\cite{DF-Lr,EK-Lr,ZFF-Lr,FDKU-Lr,BEK-Lr}. Because of small
value of this energy interval it is very sensitive to the
correlations. Different treatment of correlations lead to
significantly different values of the interval (see Table
\ref{t:Lr}). This in turn leads to large uncertainty in the value of
the polarizabilities of both states of Lr.

The value of the electric dipole transition amplitude between
$7p_{1/2}$ and $6d_{3/2}$ states of Lr in the calculations is given by
\begin{equation}
  \langle 7s^2 7p_{1/2} || \mathbf{D} || 7s^2 6d_{3/2} \rangle = 2.02
  \ {\rm a.u.}
\label{eq:7p6d}
\end{equation}
This allows us to separate the contribution due to this transition
from the rest of the sum in (\ref{eq:alpha0}) and (\ref{eq:alpha2})
and present polarizabilities in the form
\begin{eqnarray}
  \alpha_0(7p_{1/2}) = 126 + 1.35/\Delta E, \\
  \alpha_0(6d_{3/2}) = 67 - 0.677/\Delta E, \\
  \alpha_2(6d_{3/2}) = 26 + 0.677/\Delta E,
\end{eqnarray}
where all values are in atomic units and $\Delta E = E(6d_{3/2}) - E(7p_{1/2})$.
Sensitivity of the polarizabilities to the value of this energy
interval is the main source of uncertainty. The uncertainty assigned
to the polarizabilities of lawrencium (Table~{\ref{t:pol}, MBPT column) are based
on the variation of the energy interval in different calculations
(Table~\ref{t:Lr}). The uncertainties for other atoms are smaller due to 
absence of the resonance contribution. The most accurate values are those 
obtained in the all-order calculations while the difference between all-order
and MBPT results can serve as en estimation of theoretical uncertainty.

\begin{table}
\caption{Ground state scalar $\alpha_0$ and tensor $\alpha_2$ polarizabilities of Yb, Lu, Hf, No, Lr, and Rf. CI+MBPT and CI+all-order results are listed in columns labelled ``MBPT'' and ``All-order'', respectively. Last column presents the values of $\alpha_0$ from
  other sources. All numbers are in atomic units. To convert them into
$10^{-24} {\rm cm}^3$ one should divide the numbers by 6.749.}
\label{t:pol}
\begin{ruledtabular}
\begin{tabular}{llccccc} \multicolumn{2}{c}{Atom/} & \multicolumn{2}{c} {$\alpha_0$} &\multicolumn{2}{c} {$\alpha_2$}
&$\alpha_0$ \\
\multicolumn{2}{c}{State} & MBPT & All-order & MBPT &All-order &Other \\
\hline
Yb & $^1$S$_0$     &  141(6)\footnotemark[1] &141(2)\footnotemark[2] & 0 & 0 &139.3(4.9)\footnotemark[3] \\
Lu & $^2$D$_{3/2}$ &137(7)  &145 & -15(1)      &-22& 148\footnotemark[4] \\
Hf & $^3$F$_2$     &103(5)  & 97& -0.41(2)    &-0.92 & 109\footnotemark[4] \\
No & $^1$S$_0$     &112(6)  &110 &      0      & 0&\\
Lr & $^2$P$_{1/2}$ &320(80) &323 &      0     & 0 &\\
Lr & $^2$D$_{3/2}$ &-12(25) &-12 & 120(25)  &119& \\
Rf & $^3$F$_2$     &107(5)  & 115& 2.3(4)  &  8.9  & \\
\end{tabular}
\noindent \footnotetext[1]{Agrees precisely with our previous calculations,
  Ref.~\cite{DD10,DKF14}.}
\noindent \footnotetext[2] {Ref.~\cite{Yb}.}  
\noindent \footnotetext[3]{Experimental constrain, Ref.~\cite{Beloy}.}
\noindent \footnotetext[4]{Relativistic linear response calculations by
  G. D. Doolen, unpublished, cited from Ref.~\cite{Miller}.}
\end{ruledtabular}
\end{table}
Knowing the value of the electric dipole transition amplitude
(\ref{eq:7p6d}) allows us to calculate lifetime of the $6d_{3/2}$
state. It is 0.23~ms if we take the energy interval to be our theoretical value of  1555 cm$^{-1}$ (see
Table~\ref{t:Lr}). This is a long-lived metastable state. Since lawrencium atoms
are not found in nature but produced on accelerators they can probably
be produced in either of the $7p_{1/2}$ or $6d_{3/2}$ states. The
interaction with environment is very different for Lr atoms in these
two states. It is isotropic for the atoms in the $7p_{1/2}$ state and
strongly anisotropic for atoms in the $6d_{3/2}$ state. In the later
case, the polarizability is dominated by the tensor term. The total
value is positive ($\alpha \approx 100$ a.u.) for the case when total
atomic angular momentum is parallel to the electric field ($|M|=J$)
and it is negative ($\alpha \approx -160$ a.u.) for the case when total
atomic angular momentum is perpendicular to the electric field ($M=0$).\\

\section{Conclusion}

Energy levels for lowest states of superheavy elements nobelium,
lawrencium and rutherfordium as well as first few ionization
potentials and static polarizabilities  have been
calculated using the combination of the configuration interaction with
the all-order single-double methods. The accuracy of the calculations
are controlled by performing similar calculations for lighter analogs of
the elements, ytterbium, lutecium, and hafnium. These calculations provide benchmark data, critically evaluated for their accuracy, for future experimental studies.

\acknowledgments

The work was supported in part by the Australian Research Council and by USA NSF Grant No.\ PHY-1212442.

\end{document}